\documentstyle[amssymb,aps,twocolumn]{revtex}
\begin{document}
\title{Zurek-Kibble domain structures: The Dynamics of Spontaneous Vortex formation
in Annular Josephson Tunnel Junctions}
\author{R. Monaco$^{a}$\thanks{%
E-mail: roberto@sa.infn.it}, J.\ Mygind$^{b}$\thanks{%
E-mail: myg@fysik.dtu.dk} and R.\ J.\ Rivers$^{c}$\thanks{%
E-mail: r.rivers@ic.ac.uk Permanent address: Blackett Laboratory, Imperial
College, London, SW7 2BZ, U.K.}}
\address{{\it a}) Istituto di Cibernetica del C.N.R., I-80078, Pozzuoli, Italy\\
and Unita' INFM-Dipartimento di Fisica, Universita' di Salerno,\\
I-84081 Baronissi, Italy\\
b) Department of Physics, Technical University of Denmark, \\
B309, DK-2800 Lyngby, Denmark\\
{\it c}) U.K. Centre of Theoretical Physics, University of Sussex, \\
Brighton, BN1 9QJ, U.K.}
\date{\today }
\maketitle

\begin{abstract}
Phase transitions create a domain structure with defects, that has been
argued by Zurek and Kibble to depend in a characteristic way on the quench
rate. In this letter we present an experiment to measure the ZK scaling
exponent $\sigma $. Using long symmetric Josephson Tunnel Junctions, for
which the predicted index is $\sigma =0.25$, we find $\sigma =0.27\pm 0.05$.
Further, we agree with the ZK prediction for the overall normalisation.

PACS Numbers : 11.27.+d, 05.70.Fh, 11.10.Wx, 67.40.Vs
\end{abstract}

\medskip

Because phase transitions take place in a finite time, causality guarantees
that correlation lengths remain finite. Order parameter fields become
frustrated, and defects arise so as to mediate the correlated regions with
different ground states. Since defects are, in principle, observable, they
provide an excellent experimental signature for the way in which transitions
are implemented.

For condensed matter systems, which include the long annular Josephson
Tunnel Junctions (JTJs) that we shall discuss below, Zurek\cite
{zurek1,zurek2} suggested that causality alone is sufficient to determine
the initial density of defects arising in a continuous transition. In this
he paralleled proposals made by Kibble\cite{kibble1} in the context of
quantum field theory models of the early universe.

As applied to JTJs, the idea is as follows. Consider a thin linear JTJ with
critical temperature $T_{c}$, cooled through that temperature so that, if $%
T(t)$ is the temperature at time $t$, then $T(0)=T_{c}$. ${\dot{T}}%
(0)=-T_{c}/\tau _{Q}$ defines the quench time $\tau _{Q}$. Suppose, at time $%
t$, that ${c}(t)={c}(T(t))$ is the Swihart velocity\cite{Swihart,Barone},
vanishing at $t=0$, and that $\xi _{ad}(t)=\xi _{ad}(T(t))$ is the adiabatic
healing length (the Josephson length $\lambda_J(T(t))$), diverging at $t=0$.
The first time that fluxons (or Josephson vortices), the defects of linear
JTJs, can appear is at time ${\bar{t}}$, when $|{\dot \xi} _{ad}(t)|\simeq
c(t)$.

For the case in hand, ${\bar{t}}$ has the form ${\bar{t}}=\sqrt{\tau
_{Q}\tau _{0}}$, where $\tau _{0}\ll \tau _{Q}$ is the relaxation time of
the longest wavelength modes. Details are given in papers \cite{KMR,MRK} by
two of us (R.M and R.R). 
%and the reader is referred to them for further explanation.
As a result, $\tau _{Q}\gg {\bar{t}}\gg \tau _{0}$. If $\xi _{ad}(t)\sim \xi
_{0}(t/\tau _{Q})^{-\nu }$ for $t\sim 0$, where $\xi _{0}$ is simply related
to $\xi _{ad}(T=0)$, the cold fluxon size, then the initial domain size and
fluxon separation is predicted to be 
\begin{equation}
{\bar{\xi}}\sim \xi _{ad}({\bar{t}})=\xi _{0}\bigg(\frac{\tau _{Q}}{\tau _{0}%
}\bigg)^{\sigma }\gg \xi _{0},  \label{xibar}
\end{equation}
where $\sigma =\nu /2$. %This is very large on the scale of coldfluxons.
We term $\sigma $ the Zurek-Kibble (ZK) characteristic index.

The arguments are not specific to JTJs. Prior to our experiment, five other
condensed matter experiments had been performed to test the prediction (\ref
{xibar}) for the separation of defects at their production, two experiments%
\cite{grenoble,helsinki} on superfluid $^{3}He$, two\cite
{lancaster,lancaster2} on superfluid $^{4}He$, and one\cite{technion} on
high temperature superconductors (HTSC). In addition, an experiment\cite
{Roberto} on JTJs by two of us (R.M and J.M) was compatible with (\ref{xibar}%
), although it had not been performed with a test of (\ref{xibar}) in mind. 
%It was this that motivated the experiment described here.

Before discussing our new experiment a few comments are in order. The
experiments\cite{grenoble,helsinki} on superfluid $^{3}He-B$ rely on the
fact that, when it is bombarded with slow neutrons, $n+^{3}He\rightarrow
p+^{3}H+760$keV. The energy released in such a collision leads to a hot spot
which, when cooled by its environment below $T_{c}$, leaves behind a tangle
of vortices (the topological defects in this system). $\tau _{Q}$ is fixed
by the nuclear process that breaks up the $^{3}He$ atom. With only a single
data point conflating both normalisation and $\sigma $ it is not possible to
confirm the predicted value $\sigma =1/4$. However, both experiments are
highly compatible with (\ref{xibar}), agreeing to a factor of a few in each
case.

In principle, the $^{4}He$ experiments\cite{lancaster,lancaster2}, which use
a pressure quench with a varying timescale $\tau _{Q}$ to implement the
transition, allow for a more complete test. % (in
Yet again, vortices are the relevant defects. In practice, the most reliable
experiment\cite{lancaster2} sees no vortices. This is not necessarily a sign
of failure in that it has been suggested\cite{ray2} that the vortices decay
too fast to be seen. This is irrespective\cite{ray} of whether a pressure
quench, which preserves high thermal fluctuations, would of itself lead to
somewhat different predictions. In this context, the vortices seen in an
earlier $^{4}He$ experiment\cite{lancaster} were most likely an artefact of
the experimental setup.

The fifth experiment\cite{technion}, on high-$T_{C\text{ }}$superconductors,
measures total flux through a surface carried by the Abrikosov vortices. The
vortex separation of (\ref{xibar}) can be converted into a prediction for
the flux, but no flux is seen in contradiction with this prediction, despite
the phase separation that leads to the result being demonstrated elsewhere%
\cite{carmi}. In this case there is no obvious explanation of the null
result, despite later work \cite{rajantie} that takes the effect of gauge
fields into account fully.

In summary, these early experiments (including \cite{Roberto}) have either
provided {\it one} data point for (\ref{xibar}), or have been null.
Subsequently, two experiments of a very different type have been performed
that permit varying quench rates and so an estimate for $\sigma $. The most
recent\cite{pamplona} involves the B$\acute{e}$nard-Marangoni
conduction-convection transition. The defects here are not associated with
the line zeroes of an order parameter field, and the viscosity-dependent $%
\sigma $ does not match the ZK prediction, most likely for that reason. The
more relevant experiment\cite{florence} is carried out in a non-linear
optical system, with complex beam-phase the order parameter. Increasing the
light intensity (the control parameter in this case) leads to pattern
formation (defects) at a critical value. The predicted $\sigma =1/4$ is
recovered to good accuracy, but agreement with normalisation is not stated.

Our experiment, whose methodology and results we outline below, is also one
in which, by varying $\tau _{Q}$, we can measure and compare $\sigma $ with
its theoretical value, as well as confirming overall scale.

In its essence, we quench a long annular JTJ through its critical
temperature and count such fluxons or Josephson vortices as appear. In its
idealised form the annular JTJ consists of two dimensionally identical
annuli of superconductors of narrow width, superimposed upon one another,
separated by an insulating barrier in the same plane. In practice, it is
sufficient for the lower superconductor to be a planar substrate upon which
the other annulus sits. The effective theory for fluxons in JTJs\cite{Barone}
is the sine-Gordon model with respect to the field $\phi =\phi _{1}-\phi
_{2} $, the difference in the phases of the complex order parameter fields
in the separate superconductors. The Josephson vortices are then the
sine-Gordon kinks.

The JTJs in our experiment are {\it symmetric}, by which is meant that the
electrodes are made of identical superconducting material with the same
energy gaps and the same $T_{c}$. This is confirmed by seeing that a) there
is no logarithmic singularity in the voltage-current characteristic at
finite voltages and b) the temperature dependence of the critical current is
linear as $T$ approaches $T_{c}$. The outcome is that\cite{KMR,MRK} $\nu
=1/2 $. Therefore, at the time of their formation the separation of fluxons
is expected to be given by (\ref{xibar}) with $\sigma =1/4$.

In terms of the parameters of the JTJs, the Josephson length at temperature $%
T$ is\cite{Barone} 
\[
\xi_{ad}(T) =\lambda _{J}(T)=\sqrt{\frac{\hbar }{2e\mu _{0}d_{s}J_{c}(T)},} 
\]
where $J_{c}(T)$ is the critical Josephson current at temperature $T$.
Typically $\lambda _{J}(0)$ is in the $10-100\,\mu m$ range and was equal to 
$7\,\mu m$ for the sample discussed below. In Ref.\cite{MRK} $\xi _{0}$ has
been inferred as 
\[
\xi _{0}=\sqrt{\frac{\hbar }{2e\mu _{0}d_{s}\alpha J_{c}(0)}}. 
\]
The parameter $3\lesssim \alpha \lesssim 5$ is given in terms of the
superconductor gap energy $\Delta (T)$ and $T_c$. As for $\tau _{0}$, it is
given as $\tau _{0}=\xi _{0}/c_{0}$, where $c_{0}$ defines the behavior $%
c(t)=c_{0}(t/\tau _{Q})^{1/2}$of the Swihart velocity for the system near $%
T=T_{c}$. If the thickness of the two superconducting electrodes differs,
the effective thickness $d_{s}$ is the harmonic mean of the individual
thicknesses\cite{Barone}.

Our samples are high quality, $500\,\mu m$ long, $3\,\mu m$ wide, $%
Nb/Al-Al_{ox}/Nb$ JTJs fabricated on $0.5\,mm$ thick silicon substrates
using the trilayer technique (SNEAP) in which the junction is realized in
the window opened in a $SiO$ insulator layer. Details of the fabrication
process can be found in Ref.\cite{VPK}. For all samples the high quality has
been inferred by a measure of the I-V characteristic at $T=4.2\,K$ . In
fact, the subgap current $I_{sg}$ at $2\,mV$was small compared to the
current rise $\Delta I_{g}$ in the quasiparticle current at the gap voltage $%
V_{g}$, typically $\Delta I_{g}>35I_{sg}$. The gap voltage was as large as $%
V_{g}=2.76\,mV$ and the maximum critical current $I_{c}$ was larger than $%
0.55\Delta I_{g}$ for the overlap type junction. Furthermore, the
application of a strong enough external magnetic field in the barrier plane
completely suppressed any Josephson structures indicating the absence of
electrical shorts in the barrier. At an order of magnitude level, we have $%
\tau _{0}\sim 0.1\,$ps and $\xi _{0}\sim 1\,\mu $m. As a result ${\bar{\xi}}%
\sim 1\,$mm for $\tau _{Q}\sim 1\,$s. We will be more specific later.

Our JTJs have a critical temperature $T_{c}=8.95\,K$, whereas the individual
superconductors have a critical temperature of $9.1\,K$. Even at our fastest
quench the conductors are superconducting, by which is meant that the
Cooper-pair order parameter field has achieved its final magnitude, $1\,ms$
before the JTJ can develop fluxons. This is necessary for (\ref{xibar}) to
be valid without modification \cite{KMR,MRK}, since only then is $\phi $ the
relevant order parameter.

In order to vary the quenching time over the widest possible range, we have
realized the experimental setup shown schematically in Fig.1. The annular
JTJ samples are fabricated on a chip (shown endways on), mounted to a $Cu$
block by a thermally insulating teflon sheet. The entire system is enclosed
in a vacuum-tight can immersed in the liquid $He$ bath. In all cases the
heat is removed from the system by $He$ exchange gas, using a manual pump.
By varying the pressure of the gas we can modify the rate of cooling of the
sample. On the other side of the block is a $50\,\Omega $ carbon resistor,
that enables us to heat and cool the JTJs on a relatively long timescale ($%
\tau _{Q}$ from about $1$\thinspace to $10\,s$) depending on the He exchange
gas pressure inside the can. With $\sigma = 1/4$, we need to vary $\tau_Q$
by at least two orders of magnitude. This is not possible just by heating
the block alone (even if it were smaller). To extend the range of $\tau_Q$
we mounted a small, pulse-driven, surface $100\,\Omega $ resistor on the
same side. This permits much smaller thermal cycles ($0.07$\thinspace to $%
0.2\,s$). These two completely different quenching techniques provide
timescale ranges that do not overlap, leaving a gap between $0.2$ and $1s$,
that would require a third quenching technique to be filled.

The whole system is then enclosed in a $\mu $-metal shielded cryostat. The
temperature of the JTJ is monitored by measuring the junction gap voltage,
which is proportional to the known superconductor gap energy $\Delta (T(t))$%
. All quenches were taken from $10\,K$ to $4.2\,K$ through $T_{c}=8.95\,K$.
By making use of the Thouless equation\cite{thouless}, it is possible to
infer the temperature of the JTJ from the gap voltage in the range $8.2\,K$
to $4.2\,K$, extremely accurately at the upper end of the range, and with
fluctuations of a few per cent at the lower. Whether for slow or fast
cooling an excellent fit to the temperature of the JTJ in this range, for
initial temperature $T_{in}=10\,K$ and final temperature $T_{fin}=4.2\,K$,
is given by the thermal relaxation equation 
\begin{equation}
T(t)=T_{in}+(T_{in}-T_{fin})\,e^{-(t-t_{0})/\tau },  \label{T(t)}
\end{equation}
where $\tau $ is the relaxation time which sets the cooling timescale. See
Fig.2 for an example. This equation is then used to extrapolate $dT(t)/dt$
to the vicinity of $T_{C}$ and yields $\tau _{Q}\simeq 1.7\tau $. 
%More cautiously, we
%estimate errors in $\tau _{Q}$ at the $10\%$ level.

On cooling the system in this way, we expect fluxons to appear from the
inhomogeneity of $\phi $ at the transition, according to (\ref{xibar}). In
the absence of any current through the barrier or applied external magnetic
field the fluxons are in indifferent equilibrium as far as the barrier is
homogeneous and pin-hole free. (In reality, there is a small pinning effect
so that, after a short transient, the fluxons are static.) To make them
visible, and countable, we apply a bias current, whereby they move as
magnetic dipoles under the resulting Lorentz force, at speed less than the
Swihart velocity. According to Josephson theory $N$ travelling fluxons (and
antifluxons) with speed $v$ develop a voltage $V=N\Phi _{0}v/C$ across the
junction, where $\Phi _{0}=h/2e$ is the flux in a (Josephson) fluxon and $C$
is the annulus circumference. This voltage can be measured, and the fluxon
number determined. A detailed description of the experiment and the data
will be given elsewhere\cite{MMR}.

Many samples have been measured, but only one had such a large critical
current density (and sufficiently small $\lambda_J(T)$) that only 3000
thermal cycles were enough to get reliable statistics, and it is this sample
that we shall discuss now. However, we stress that, within the less good
statistics of the other samples, none gave results that were incompatible
with (\ref{xibar}).

The symmetric annular JTJ with which the experiment was performed had a
circumference $C=500\,\mu m$, and width $\Delta r=4\,\mu m$. The effective
superconductor thickness was $d_{s}\approx 250$ $nm$. At the final
temperature $T_{fin}=4.2\,K$, the critical current density was $%
J_{c}(T_{fin})=3050\,A/cm^{2}$, the quality factor was $V_{m}=49mV$and the
Josephson length was $\lambda _{J}(T_{fin})=6.9\,\mu m$. The velocity $c_{0}$
is calculated to be $c_{0}\approx 2.2\times 10^{7}$\thinspace $m/s$. From
this, we infer that $\xi _{0}\approx 3.8\,\mu m$ and $\tau _{0}\approx
0.17\,ps$. It follows that ${\bar{\xi}}\approx 5.9\,mm$ for $\tau _{Q}=1\,s$%
, several times the circumference of the JTJ.

As a result, the likelihood of finding a single fluxon is small. We estimate
the probability of finding a fluxon in a single quench to be 
\begin{equation}
P_1 \simeq \frac{C}{\bar\xi} = \frac{C}{\xi_0}\bigg(\frac{\tau_Q}{\tau_0}%
\bigg)^{-\sigma},  \label{P1}
\end{equation}
where, from (\ref{xibar}), $\sigma = 0.25$.

In Fig.3 we show a log-log plot for the measured $P_{1}$ as a function of $%
\tau _{Q}$, changed by varying the exchange gas pressure and by using both
the fast and slow quenching techniques. Each data point corresponds to 300
quenches. Although the data does not distinguish between a single fluxon and
a fluxon plus an antifluxon pair, the likelihood of the latter is
sufficiently small that it can be ignored. Similarly, the data does not
distinguish between no fluxons and a fluxon-antifluxon pair, with similar
errors. We observe that the points are quite scattered, meaning that the
data are statistically poor. Further, for the reasons given earlier there is
a gap between fast and slow quenches, with the possibility for systematic
error. Nonetheless, we have clear evidence that i) the trapping of a fluxon
occurs on a purely statistical basis for identical conditions of each
thermal cycle and ii) the probability to trap one fluxon is larger when the
transition is performed at a faster speed (smaller quenching time) in
accordance with the causality principle. This complements our qualitative
results from other samples with smaller $J_c(0)$ (for which the statistics
is too poor to permit a fit to (\ref{xibar})) that, at fixed $\tau_Q$, the
probability of finding a fluxon decreases with increasing $\xi_0$.

Regardless of the data spread, to test (\ref{P1}) we attempted to fit the
data with an allometric function $P_{1}=a\,\tau _{Q}^{-b}$, with $a$ and $b$
being free fitting parameters. We found, for the coefficient $a$, the best
fitting value of $0.1\pm 10\%$ (taking $\tau _{Q}$ in seconds). This is in
excellent agreement with the predicted value of $C\tau _{0}^{1/4}/\xi
_{0}\approx 0.08\,s^{1/4}$, given the fact that we only expect agreement in
overall normalization to somewhat better than an order of magnitude. After
the failure of the experiments discussed in \cite
{lancaster,lancaster2,technion} to find (reliable) defects at expected
densities, if at all, this experiment shows that the ZK estimate remains
sensible. Further, the best fitting curve, shown by the solid line in Fig.3,
has a slope $b=0.27\pm 0.05$, in remarkable agreement with its predicted
value of $0.25$.

The ZK scenario needs further testing with JTJs for which there is a greater
likelihood of observing fluxons. In \cite{KMR} and \cite{MRK} we observed
that this is the case for significantly non-symmetric JTJs, for which the
value of $\sigma $ is $\sigma =1/7$; a further experiment, with markedly
non-symmetric JTJs, is being planned.

\smallskip

The authors thank L. Filippenko for the sample fabrication and V.P.
Koshelets for usefull discussions. R.R. thanks the University of Salerno for
hospitality. This work is, in part, supported by the COSLAB programme of the
European Science Foundation, the Danish Research Council, and the Hartmann
Foundation.

\section{\bf Figure Captions}

%\addvspace{3truecm}

\begin{itemize}
\item[Fig.1]  Sketch (dimensions are not to scale) of the cryogenic insert
developed to perform the junction thermal cycles with a time scale changing
over a broad range.The JTJs are fabricated within the chip (shown endways),
to which the surface mounted resistor (SMR) is attached. The whole is
surrounded with liquid He.

\item[Fig.2]  Time dependence of the junction temperature during a ''slow''
thermal cycle from 10 K to 4.2 K, as inferred from the junction gap voltage.
Only below the horizontal dashed line does the continuous curve describe the
temperature. The thick dashed line is the best fitting curve to (\ref{T(t)}%
), to be extrapolated to $T_{c}$.

\item[Fig.3]  Log-log plot of the measured probability $P_{1}$ to trap one
fluxon versus the quenching time $\tau _{Q}$. Each point corresponds to 300
thermal cycles. The solid line (slope $b=0.27$) is the best linear fit, in
good agreement with the $0.25$ value expected for symmetric JTJs. Errors in $%
\tau _{Q}$ are a few per cent, and systematic errors in $P_{1}$ due to the
neglect of fluxon-antifluxon pairs are $O(P_{1}^{2})$, again a few percent.
\end{itemize}

\end{document}